\documentclass{article}%
\usepackage{amsfonts}
\usepackage{amsmath}
\usepackage{amssymb}
\usepackage{graphicx}%
\providecommand{\U}[1]{\protect\rule{.1in}{.1in}}

\begin{document}

\title{Exact Renormalization Groups as a form of Entropic Dynamics\thanks{Presented
at MaxEnt 2017, the 37th International Workshop on Bayesian Inference and
Maximum Entropy Methods in Science and Engineering (July 9-14, 2017, Jarinu,
Brazil). }}
\author{Pedro Pessoa and Ariel Caticha\\{\small Department of Physics, University at Albany--SUNY, Albany, NY 12222,
USA} }
\date{}
\maketitle

\begin{abstract}
The Renormalization Group (RG) is a set of methods that have been instrumental
in tackling problems involving an infinite number of degrees of freedom. What
all these methods have in common -- which is what explains their success -- is
that they allow a systematic search for those degrees of freedom that happen
to be relevant to the phenomena in question. In the standard approaches the RG
transformations are implemented by either coarse graining or by changes of
variables. When these transformations are infinitesimal the formalism can be
described as a continuous dynamical flow in a fictitious time parameter. It is
generally the case that these exact RG equations are functional diffusion
equations. In this paper we show that the exact RG equations can be derived
using entropic methods. The RG flow is then described as a form of entropic
dynamics of field configurations. Although equivalent to other versions of the
RG, in this approach the RG transformations receive a purely inferential
interpretation that establishes a clear link to information theory.

\end{abstract}

\section{Introduction}

The Renormalization Group (RG) is a collection of techniques designed for
tackling problems that involve an infinite number of coupled degrees of
freedom. The range of applications is enormous, it includes quantum field
theory, the statistical mechanics of critical phenomena, turbulence, and many
others. Ever since the work of Wilson (see \emph{e.g. }\cite{Wilson Kogut
1974}\cite{Wilson 1983}) it has been clear that the various RGs succeed
because they provide a systematic procedure to construct an effective theory
for those variables that are most relevant to the problem at hand. For
example, in Wilson's approach to critical phenomena the procedure consists in
gradually integrating out the degrees of freedom with short wavelengths to
obtain an effective Hamiltonian for the long wavelengths that are empirically
relevant \cite{Wilson Kogut 1974}.

The RG transformations are implemented by either eliminating degrees of
freedom through coarse graining, or by changes of variables, or by a
combination of the two. The result is that the RG transformations generate a
continuous flow in the statistical manifold of Gibbs distributions. One
crucial early insight \cite{Wilson Kogut 1974}\cite{Wegner Houghton 1973} was
that infinitesimal RG transformations could be implemented
exactly.\footnote{This formalism is now variously known as the exact RG, the
functional RG, and the non-perturbative RG.} This has both conceptual and
computational advantages. On the conceptual side, for example, the work of
Polchinski \cite{Polchinski 1984} used an exact RG as a method to prove
renormalizability in quantum field theory. On the computational side, the
exact RG was extensively exploited by C. Wetterich \cite{Wetterich 1991} and
coauthors --- the effective average action method --- in statistical
mechanics, in Yang-Mills theory \cite{Reuter Wetterich 1994}, and in gravity
\cite{Reuter 1998}. More recently, also on the computational side, the work of
N. Caticha and collaborators points in the direction of deploying RG
techniques for data analysis \cite{NCaticha et al 2017}.

Another crucial contribution was Wegner's realization that the elimination of
degrees of freedom is not strictly necessary, that an appropriate change of
variables could effectively accomplish the same task \cite{Wegner 1974}. The
precise form of those changes of variables have been elaborated by a number of
authors \cite{Caticha 1984}-\cite{Bervillier 2014}. In \cite{Caticha 1984} the
reason why RGs are useful is particularly clear: the changes of variables are
such that a classical or saddle-point approximation becomes more accurate,
asymptotically approaching the exact result, and therefore offering a way to
reach beyond the limitations of perturbation theory.\footnote{For additional
references see the excellent reviews \cite{Morris 1994}-\cite{Nagy 2012}.}

In this paper we develop a new approach to the exact RG derived as an
application of entropic methods of inference --- and entropic renormalization
group.\footnote{The principle of maximum entropy as a method for inference can
be traced to the pioneering work of E. T. Jaynes \cite{Jaynes 1957}%
-\cite{Jaynes 2003}. For a pedagogical overview of Bayesian and entropic
inference and further references see \cite{Caticha 2012}.} The motivation is
two-fold. First, although it is equivalent to other versions of the exact RG,
in this approach the RG transformations receive a purely inferential
interpretation that establishes a clear link to information theory. Second, it
turns out that the RG flow is described as a form of Entropic Dynamics (ED).
ED had previously been deployed to derive quantum theory as a form of
inference both for particles (see, \emph{e.g.}, \cite{Caticha 2015}%
\cite{Caticha 2017}) and for fields \cite{Ipek Caticha 2014}. The formulation
of an ED version of RG presented here is a first step towards establishing a
closer link between RG techniques and the foundations of quantum field theory.
The natural expectation is that this will lead to further insights into
Yang-Mills and gravity theory.

In sections 2 and 3, we establish notation and give a brief review of the RG
as an exact change of variables. The derivation of the RG as a form of
Entropic Dynamics is given in section 4.

\section{Some background and notation}

Our subject is the statistical mechanics of a scalar field $\phi(x)=\phi_{x}$
in $d$ spatial dimensions; such a field configuration can be represented as a
point $\phi$ in an $\infty$-dimensional configuration space $\mathcal{C}$. The
Fourier components are denoted%
\begin{equation}
\phi_{q}=\int dx\,\phi_{x}e^{iq\cdot x}~,
\end{equation}
where $dx=d^{d}x$. In thermal equilibrium the probability distribution of
$\phi_{x}$ is of the Gibbs form,
\begin{equation}
\rho\lbrack\phi]=\frac{1}{Z}e^{-H[\phi]}\quad\text{where}\quad Z=\int\left(
{\textstyle\prod\nolimits_{q}}
d\phi_{q}\right)  \,e^{-H[\phi]} \label{Gibbs}%
\end{equation}
is the partition function, and a factor $\beta=1/kT$ has been absorbed into
the Hamiltonian $H$.

In this section, for simplicity, we describe the paradigmatic example of a
sharp-cutoff RG. The sequence of RG\ transformations generates a trajectory of
effective Hamiltonians $H_{\tau}$ labeled by a parameter $\tau$. Once the
integration over all $\phi_{q}$s with $q$ higher than a certain cutoff
$\Lambda$ has been performed, the partition function takes the form
\begin{equation}
Z=\int\left(
{\textstyle\prod\nolimits_{q<\Lambda}}
d\phi_{q}\right)  \,e^{-H_{\tau}[\phi]}\quad\text{where}\quad e^{-H_{\tau
}[\phi]}=\int\left(
{\textstyle\prod\nolimits_{q>\Lambda}}
d\phi_{q}\right)  \,e^{-H[\phi]}~.
\end{equation}
The infinitesimal RG transformation requires two steps. The first involves
integrating out those wavelengths in the narrow shell with $\Lambda
e^{-\delta\tau}<q<\Lambda$ leading to
\begin{equation}
Z=\int\left(
{\textstyle\prod\nolimits_{q<\Lambda e^{-\delta\tau}}}
d\phi_{q}\right)  \,e^{-H_{\tau+\delta\tau}^{\prime}[\phi]}~,
\end{equation}
where
\begin{equation}
e^{-H_{\tau+\delta\tau}^{\prime}[\phi]}=\int\left(
{\textstyle\prod\nolimits_{\Lambda e^{-\delta\tau}<q<\Lambda}}
d\phi_{q}\right)  \,e^{-H_{\tau}[\phi]}~.
\end{equation}
Since this is an infinitesimal transformation it can be carried out exactly
\cite{Wegner Houghton 1973}\cite{Caticha 1984}. The result is
\begin{equation}
H_{\tau+\delta\tau}^{\prime}-H_{\tau}=(2\pi)^{d}\Lambda^{d-2}d\tau\int
d\Omega_{d}\left[  \frac{\delta^{2}H_{\tau}}{\delta\phi_{q}\delta\phi_{-q}%
}-\frac{\delta H_{\tau}}{\delta\phi_{q}}\frac{\delta H_{\tau}}{\delta\phi
_{-q}}\right]  \label{sharp cutoff}%
\end{equation}
where $q^{2}=$ $\Lambda^{2}$ and $d\Omega_{d}\,$\ is the element of solid
angle in $d$ dimensions. The typical RG transformation includes a second step
in which momenta and fields suitably re-scaled to yield $H_{\tau+\delta\tau}$.
The momenta are scaled by $q\rightarrow qe^{\delta\tau}$ so that throughout
the RG\ flow the new momenta always span the same constant range $(0,\Lambda
)$. The rescaling of the fields is
\begin{equation}
\phi_{x}\rightarrow\phi_{x^{\prime}}^{\prime}=e^{d_{\phi}\delta\tau}\phi
_{x}\quad\text{or}\quad\phi_{q}\rightarrow\phi_{q^{\prime}}^{\prime
}=e^{(d_{\phi}-d)\delta\tau}\phi_{q}~, \label{scaling a}%
\end{equation}
where the field scale dimension $d_{\phi}=d/2-1+\gamma_{\phi}$ includes the
$\gamma_{\phi}$ correction --- the anomalous dimension --- needed for the
trajectory to flow towards a fixed point $H_{\infty}$ as $\tau\rightarrow
\infty$.

\section{The RG as a change of variables}

One advantage of expressing the partition function as an integral is that we
can easily study the effects induced by transformations of the dynamical
variables. This allows us to explore the idea that the RG is a technique that
selects the relevant variables as they transform through different scales.
Generalizing beyond the sharp cutoff case discussed in the previous section,
the partition function at some stage $\tau$ of the RG flow can, in general, be
written as
\begin{equation}
Z=\int D\phi\,e^{-H_{\tau}[\phi]}\quad\text{where}\quad D\phi=%
{\textstyle\prod\nolimits_{q}}
d\phi_{q}~, \label{Z}%
\end{equation}
with no limitations on the range of $q.$ As $\tau\rightarrow-\infty$, the
effective Hamiltonian tends to the bare Hamiltonian in (\ref{Gibbs}),
$H_{\tau}\rightarrow H_{-\infty}=H$. Consider an infinitesimal change of
variables,
\begin{equation}
\phi_{q}\rightarrow\phi_{q}^{\prime}=\phi_{q}-\delta\tau\,\eta_{\tau q}%
[\phi]~, \label{ch var a}%
\end{equation}
where $\eta_{\tau q}[\phi]$ is some sufficiently well-behaved functional of
$\phi$ and a function of $q$. Then eq.(\ref{Z}) becomes
\begin{equation}
Z=\int D\phi\,\left[  1-\delta\tau\int dq\frac{\delta\eta_{\tau q}[\phi
]}{\delta\phi_{q}}\right]  \exp-\left[  H_{\tau}[\phi]-\delta\tau\int
dq\frac{\delta H_{\tau}}{\delta\phi_{q}}\eta_{\tau q}[\phi]\right]  \,,
\end{equation}
where $dq=d^{d}q/(2\pi)^{d}$. This leads to
\begin{equation}
Z=\int D\phi\,\exp-H_{\tau+\delta\tau}[\phi]~,
\end{equation}
\ where%
\begin{equation}
H_{\tau+\delta\tau}[\phi]=H_{\tau}[\phi]-\delta\tau\int dq\left[  \frac{\delta
H_{\tau}}{\delta\phi_{q}}\eta_{\tau q}[\phi]-\frac{\delta\eta_{\tau q}[\phi
]}{\delta\phi_{q}}\right]  ~. \label{RG a}%
\end{equation}

As discussed in \cite{Caticha 1984}, the choice of $\eta_{\tau}$ that
reproduces an RG transformation (see, \emph{e.g.}$,$\emph{ }%
eq.(\ref{sharp cutoff})) is $\eta_{\tau q}[\phi]\sim$ $\delta H_{\tau}%
/\delta\phi_{-q}$. The effect of integrating out short wavelengths as opposed
to long wavelengths is achieved by an appropriate $q$-dependent
proportionality constant $f_{q}$. Typically we want some positive $f_{q}$ that
leaves long wavelengths unmodified while effectively integrating out the short
wavelengths. A suitable choice is, for example, $f_{q}\sim q^{2}/\Lambda^{2}$,
so that $f_{q}$ is small for $q\ll\Lambda$, and $f_{q}$ is large for
$q\gg\Lambda$, where $\Lambda$ is some reference momentum. The complete
RG\ transformation also involves an additional scaling of momenta
$q\rightarrow qe^{\delta\tau}$ and fields, eq.(\ref{scaling a}). The full
change of variables is
\begin{equation}
\eta_{\tau q}=\,f_{q}\frac{\delta H_{\tau}}{\delta\phi_{-q}}+\zeta_{q}\phi
_{q}\quad\text{where}\quad\zeta_{q}=d-d_{\phi}+q\cdot\frac{\partial}{\partial
q}~. \label{ch var b}%
\end{equation}
The corresponding exact RG equation is
\begin{equation}
\frac{\partial}{\partial\tau}H_{\tau}=\int dq\,\left[  f_{q}\left(
\frac{\delta^{2}H_{\tau}}{\delta\phi_{q}\delta\phi_{-q}}-\frac{\delta H_{\tau
}}{\delta\phi_{q}}\frac{\delta H_{\tau}}{\delta\phi_{-q}}\right)
+\frac{\delta H_{\tau}}{\delta\phi_{q}}\zeta_{q}\phi_{q}\right]  ~.
\label{RG b}%
\end{equation}
It turns out that observable quantities such as critical exponents are
independent of the particular choice of $f_{q}$. For later convenience we
rewrite (\ref{RG a}) as an equation for $\rho_{\tau}=e^{-H_{\tau}[\phi]}/Z$.
The result is remarkably simple,
\begin{equation}
\frac{\partial}{\partial\tau}\rho_{\tau}=-\int dq\,\frac{\delta}{\delta
\phi_{q}}\left(  \rho_{\tau}\eta_{\tau q}\right)  ~. \label{RG c}%
\end{equation}

\section{The Entropic Renormalization Group}

Next we derive the RG evolution as a form of entropic dynamics. (For the
related ED of \emph{quantum} scalar fields see \cite{Ipek Caticha 2014}.) We
consider a generic probability distribution $\rho_{\tau}[\phi]$ and we wish to
study how it flows as a function of the parameter $\tau$.

The basic \textquotedblleft dynamical\textquotedblright\ assumption is that
under the RG flow the fields follow continuous trajectories. This means that a
finite transformation can be analyzed as a sequence of infinitesimally short
steps and allows us to focus our attention on infinitesimal RG transformations.

Given that a certain field configuration $\phi$ is transformed into a
neighboring one $\phi^{\prime}$, we ask, what can we expect $\phi^{\prime}$ to
be? It is common practice to define a coarse graining transformation that
allows one to calculate $\phi^{\prime}$ from the given $\phi$. Such RGs lead
to a \emph{deterministic} flow. In contrast, the essence of an entropic
dynamics is that the information about the new $\phi^{\prime}$ is very limited
and the goal is to determine a transition probability $P[\phi^{\prime}|\phi]$.
Thus, the \emph{entropic RG }leads to an inherently \emph{indeterministic} dynamics.

The transition probability $P[\phi^{\prime}|\phi]$ is assigned by maximizing
the entropy,
\begin{equation}
S[P;Q]=-\int D\phi^{\prime}\,P[\phi^{\prime}|\phi]\log\frac{P[\phi^{\prime
}|\phi]}{Q[\phi^{\prime}|\phi]}~, \label{entropy}%
\end{equation}
relative to the prior $Q[\phi^{\prime}|\phi]$, and subject to appropriate
constraints. The choice of the logarithmic entropy, as opposed to Renyi or
Tsallis entropies, is significant. The RG is a method to predict the physical
correlations between long wavelength fields; it is essential that the method
of inference itself do not contaminate the analysis by introducing unwarranted correlations.

\paragraph*{The prior--}

We adopt a prior that incorporates the information that the fields change by
infinitesimal amounts but is otherwise very uninformative. We want a prior
that does not introduce unwarranted correlations while reflecting the basic
rotational and translational symmetry of $d$-dimensional space --- a field
degree of freedom $\phi_{x}$ located at $x$ is not in any way different from
another $\phi_{x^{\prime}}$ at $x^{\prime}$. Such a prior is given by a
product of Gaussians,
\begin{equation}
Q(\phi^{\prime}|\phi)\propto\exp-\frac{1}{2\Delta\tau}\int dq\,\frac{1}%
{2f_{q}}\Delta\phi_{q}\Delta\phi_{-q}~, \label{prior}%
\end{equation}
where $\Delta\phi_{q}=\phi_{q}^{\prime}-\phi_{q}$, and the various factors of
2 are chosen for later convenience.\footnote{See eq.(\ref{curr v b}) below.
The units of $\tau$ are such that the exponent in (\ref{prior}) is
dimensionless.} The crucial factor $1/f_{q}$, see eq.(\ref{ch var b}),
enforces a different treatment for different scales; it implements the basic
idea that field components with long wavelengths remain unchanged. The limit
of infinitesimally short steps will be eventually implemented by taking
$\Delta\tau\rightarrow0$.

\paragraph*{The constraint--}

The possibility of directionality in the dynamical flow is introduced through
a constraint involving a drift potential $\Omega\lbrack\phi]$. The constraint
is that the expected value of the change of $\Omega$ in the short step
$\Delta\phi_{x}$ is some small value $\kappa$,
\begin{equation}
\left\langle \Delta\Omega\right\rangle _{P}=\kappa~. \label{constraint a}%
\end{equation}
The specific form of the drift potential $\Omega$ that implements the
rescaling of fields, and the numerical value $\kappa$ will be determined
below. This constraint can be written as
\begin{equation}
\left\langle \int dx\,\frac{\delta\Omega}{\delta\phi_{x}}\Delta\phi
_{x}\right\rangle _{P}=\int D\phi^{\prime}\,P[\phi^{\prime}|\phi]\left(  \int
dq\,\frac{\delta\Omega}{\delta\phi_{q}}\Delta\phi_{q}\right)  =\kappa~.
\label{constraint b}%
\end{equation}

\paragraph*{The transition probability--}

The distribution $P[\phi^{\prime}|\phi]$ that maximizes $S[P;Q]$ subject to
(\ref{constraint b}) and normalization is
\begin{equation}
P[\phi^{\prime}|\phi]\propto\exp-\int dq\,\left[  \frac{1}{4f_{q}\Delta\tau
}\Delta\phi_{q}\Delta\phi_{-q}-\frac{\delta\Omega}{\delta\phi_{q}}\Delta
\phi_{q}\right]  \label{P a}%
\end{equation}
where the Lagrange multiplier has been absorbed into $\Omega$. The transition
probability (\ref{P a}) is a Gaussian, more conveniently written as
\begin{equation}
P[\phi^{\prime}|\phi]=\frac{1}{Z}\exp-\frac{1}{2\Delta\tau}\int dq\,\frac
{1}{2f_{q}}\left(  \Delta\phi_{q}-\frac{\delta\Omega}{\delta\phi_{-q}}%
2f_{q}\Delta\tau\right)  \left(  \Delta\phi_{-q}-\frac{\delta\Omega}%
{\delta\phi_{q}}2f_{q}\Delta\tau\right)  . \label{P b}%
\end{equation}
This ED is a standard Wiener process. A generic step can be written as the sum
of a drift and a fluctuation, $\Delta\phi_{q}=\langle\Delta\phi_{q}%
\rangle+\Delta w_{q}$, such that
\begin{equation}
\langle\Delta\phi_{q}\rangle=\frac{\delta\Omega}{\delta\phi_{-q}}2f_{q}%
\Delta\tau\,,\quad\langle\Delta w_{q}\rangle=0\,,\quad\text{and}\quad
\langle\Delta w_{q}\Delta w_{-q^{\prime}}\rangle=2f_{q}\Delta\tau
\,\delta_{qq^{\prime}}~.
\end{equation}

\paragraph*{Entropic dynamics in integral form--}

The dynamics induced by $P[\phi^{\prime}|\phi]$ follows from the rules of
probability theory applied to the joint probability of two successive
configurations $\phi$ and $\phi^{\prime}$. Marginalizing $\rho\lbrack
\phi^{\prime},\phi]$,
\begin{equation}
\int D\phi\,\rho\lbrack\phi^{\prime},\phi]=\int D\phi\,P[\phi^{\prime}%
|\phi]\rho_{\tau}[\phi]=\rho_{\tau+\delta\tau}[\phi^{\prime}]~. \label{CK}%
\end{equation}
This is the ED equation of evolution. It describes a coarse-graining and a
drift, but notice that what is being coarse-grained here is the distribution
$\rho_{\tau}[\phi]$ and not the field configuration $\phi$ itself. Notice also
that eq.(\ref{CK}) is of the form of a Chapman-Kolmogorov equation but there
is a subtle difference in that eq.(\ref{CK}) is not meant to describe a
Markovian process that occurs in an already existing \textquotedblleft
physical\textquotedblright\ background time. Here there is no pre-existing
background time; the \textquotedblleft RG time $\tau$\textquotedblright\ is
being created by the entropic dynamics itself in such a way that, given the
\textquotedblleft present\textquotedblright\ $\rho_{\tau}$, the
\textquotedblleft future\textquotedblright\ $\rho_{\tau+\delta\tau}$ is
statistically independent of the \textquotedblleft past\textquotedblright%
\ $\rho_{\tau-\delta\tau}$.

\paragraph*{The arrow of RG time--}

Eq.(\ref{CK}) is strongly directional: $\rho_{\tau}[\phi]$\ is prior and
$\rho_{\tau+\delta\tau}[\phi^{\prime}]$\ is posterior. Applying the rules of
ED to $\rho_{\tau+\delta\tau}[\phi^{\prime}]$ leads forward to $\rho
_{\tau+2\delta\tau}[\phi^{\prime\prime}]$; they do not lead back to
$\rho_{\tau}[\phi]$. Granted, the rules of probability theory also allow us to
construct a time-reversed evolution,
\begin{equation}
\int D\phi^{\prime}\,P[\phi|\phi^{\prime}]\rho_{\tau+\delta\tau}[\phi^{\prime
}]=\rho_{\tau}[\phi]~\,,
\end{equation}
but $P[\phi|\phi^{\prime}]$ is a very different object related to
$P[\phi^{\prime}|\phi]$ by Bayes' theorem,
\begin{equation}
P[\phi|\phi^{\prime}]=\frac{\rho_{\tau}[\phi]}{\rho_{\tau+\delta\tau}%
[\phi^{\prime}]}P[\phi|\phi^{\prime}]~. \label{Bayes thm}%
\end{equation}
Thus, the asymmetry between priors and posteriors leads to an asymmetry
between the inferential past and the inferential future: if $P[\phi^{\prime
}|\phi]$ is a Gaussian derived from the maximum entropy method, then the
time-reversed $P[\phi|\phi^{\prime}]$ is obtained from Bayes' theorem and is
not Gaussian in general.

\paragraph*{Entropic dynamics in differential form--}

The ED described by (\ref{CK}) can be written as a functional differential
equation of the Fokker-Planck type,
\begin{equation}
\frac{\partial}{\partial\tau}\rho_{\tau}=-\int dq\,\frac{\delta}{\delta
\phi_{q}}\left(  \rho_{\tau}v_{q}\right)  ~, \label{FP}%
\end{equation}
where $v_{q}[\phi]$ is the $q$-component of the \textquotedblleft
current\textquotedblright\ velocity with which probabilities flow in the
$\infty$-dimensional space $\mathcal{C}$.\footnote{For algebraic details in
finite dimensions see (\cite{Caticha 2012}). The combination $%
{\textstyle\int}
dq\,\delta/\delta\phi_{q}$ is the functional equivalent of the divergence
operator.} The current velocity $v_{q}$ is the sum of two contributions, a
drift and an osmotic component
\begin{equation}
v_{q}[\phi]=b_{q}[\phi]+u_{q}[\phi]=2f_{q}\frac{\delta\Omega}{\delta\phi_{-q}%
}-f_{q}\frac{\delta\log\rho_{\tau}}{\delta\phi_{-q}} \label{curr v a}%
\end{equation}
where the first and second terms are respectively called the drift and osmotic velocities.

\paragraph*{Equivalence with the RG change of variables--}

So far we discussed the ED evolution, eq.(\ref{FP}), of a generic distribution
$\rho_{\tau}[\phi]$ in a fictitious time $\tau$. To make contact with the RG
evolution, we set $\rho_{\tau}=e^{-H_{\tau}[\phi]}/Z$ with initial condition
$H_{\tau}\rightarrow H$ (the bare Hamiltonian) as $\tau\rightarrow-\infty$,
and with $Z$ independent of $\tau$. Then the current velocity (\ref{curr v a})
is
\begin{equation}
v_{q}=f_{q}\frac{\delta H_{\tau}}{\delta\phi_{-q}}+2f_{q}\frac{\delta\Omega
}{\delta\phi_{-q}}~. \label{curr v b}%
\end{equation}
Comparing eq.(\ref{FP}) with (\ref{RG c}), which amounts to comparing
(\ref{curr v b}) with (\ref{ch var b}), shows that the ED evolution is
identical with the RG evolution provided we choose a drift potential $\Omega$
such that
\begin{equation}
2f_{q}\frac{\delta\Omega}{\delta\phi_{-q}}=\zeta_{q}\phi_{q}~.
\label{grad Omega}%
\end{equation}
The solution to this functional differential equation for $\Omega\lbrack\phi]$
is some functional that is quadratic and possibly non-local in the fields.
Fortunately, however, an explicit solution is not needed. None of the basic ED
equations --- the constraint (\ref{constraint b}), the transition probability
(\ref{P b}), and the RG\ equation (\ref{FP}) with (\ref{curr v b}) --- require
knowledge of $\Omega$; we only need to know its gradient, eq.(\ref{grad Omega}).

\section{Final remarks}

To summarize our conclusions: The evolution of probability distributions under
exact RG transformations can be formulated as a form of entropic dynamics.
This establishes a clear link between the RG and information theory. This is
not totally unexpected since the goal of the RG\ method is to select variables
that best capture the relevant information about long distance behavior, while
on the other hand, entropic methods are designed for the optimal manipulation
of information.\bigskip\ 

\noindent\textbf{Acknowledgments:} We would like to thank N\'{e}stor Caticha
for many stimulating discussions.

\end{document}